\documentclass{emulateapj}
\usepackage{apjfonts, natbib, draftcopy}

\newcommand{\be}{\begin{equation}}
\newcommand{\ee}{\end{equation}}

\newcommand{\ba}{\begin{eqnarray}}
\newcommand{\ea}{\end{eqnarray}}

\newcommand{\bp}{{\boldmath{p}}}

\shorttitle{Cosmological Parameters from QUaD}
\shortauthors{QUaD collaboration}

\begin{document}

\slugcomment{Submitted to ApJ}

\title{Cosmological Parameters from the QUaD CMB polarization experiment}

\author{
  QUaD collaboration
  --
  P.\,G.\,Castro\altaffilmark{1,2},
  P.\,Ade\altaffilmark{3},
  J.\,Bock\altaffilmark{4,5},
  M.\,Bowden\altaffilmark{3,6},
  M.\,L.\,Brown\altaffilmark{1,8},
  G.\,Cahill\altaffilmark{9},
  S.\,Church\altaffilmark{6},
  T.\,Culverhouse\altaffilmark{7},
  R.\,B.\,Friedman\altaffilmark{7},
  K.\,Ganga\altaffilmark{10},
  W.\,K.\,Gear\altaffilmark{3},
  S.\,Gupta\altaffilmark{3},
  J.\,Hinderks\altaffilmark{6,11},
  J.\,Kovac\altaffilmark{5},
  A.\,E.\,Lange\altaffilmark{5},
  E.\,Leitch\altaffilmark{4,5},
  S.\,J.\,Melhuish\altaffilmark{12},
  Y.\,Memari\altaffilmark{1},
  J.\,A.\,Murphy\altaffilmark{9},
  A.\,Orlando\altaffilmark{3,5}
  C.\,Pryke\altaffilmark{7},
  R.\,Schwarz\altaffilmark{7},
  C.\,O'\,Sullivan\altaffilmark{9},
  L.\,Piccirillo\altaffilmark{12},
  N.\,Rajguru\altaffilmark{3,13},
  B.\,Rusholme\altaffilmark{6},
  A.\,N.\,Taylor\altaffilmark{1},
  K.\,L.\,Thompson\altaffilmark{6},
  A.\,H.\,Turner\altaffilmark{3},
  E.\,Y.\,S.\,Wu\altaffilmark{6}
  and
  M.\,Zemcov\altaffilmark{3,4,5}
}

\altaffiltext{1}{Scottish Universities Physics Alliance (SUPA),
   Institute for Astronomy, University of Edinburgh,
  Royal Observatory, Blackford Hill, Edinburgh EH9 3HJ, UK.}
  \altaffiltext{2}{{\em Current address}: CENTRA, Departamento de F\'{\i}sica,
  Edif\'{\i}cio Ci\^{e}ncia,
  Instituto Superior T\'ecnico,
  Av. Rovisco Pais 1, 1049-001 Lisboa, Portugal.}
\altaffiltext{3}{School of Physics and Astronomy, Cardiff University,
  Queen's Buildings, The Parade, Cardiff CF24 3AA, UK.}
\altaffiltext{4}{Jet Propulsion Laboratory, 4800 Oak Grove Dr.,
  Pasadena, CA 91109, USA.}
\altaffiltext{5}{California Institute of Technology, Pasadena, CA
  91125, USA.}
\altaffiltext{6}{Kavli Institute for Particle Astrophysics and
Cosmology and Department of Physics, Stanford University,
382 Via Pueblo Mall, Stanford, CA 94305, USA.}
\altaffiltext{7}{Kavli Institute for Cosmological Physics,
  Department of Astronomy \& Astrophysics, Enrico Fermi Institute, University of Chicago,
  5640 South Ellis Avenue, Chicago, IL 60637, USA.}
\altaffiltext{8}{{\em Current address}: Cavendish Laboratory,
  University of Cambridge, J.J. Thomson Avenue, Cambridge CB3 OHE, UK.}
\altaffiltext{9}{Department of Experimental Physics,
  National University of Ireland Maynooth, Maynooth, Co. Kildare,
  Ireland.}
\altaffiltext{10}{APC UMR 7164 (Univ. Paris Diderot-Paris 7 - CNRS - CEA - Obs. de Paris),
  10, rue Alice Domon et L\'eonie Duquet, 75205 Paris Cedex 13, France.}
\altaffiltext{11}{{\em Current address}: NASA Goddard Space Flight
  Center, 8800 Greenbelt Road, Greenbelt, Maryland 20771, USA.}
\altaffiltext{12}{School of Physics and Astronomy, University of
  Manchester, Manchester M13 9PL, UK.}
\altaffiltext{13}{{\em Current address}: Department of Physics and Astronomy, University
  College London, Gower Street, London WC1E 6BT, UK.}


\begin{abstract}

In this paper we present a parameter estimation analysis
of the polarization and temperature power spectra from
the second and third season of observations with the QUaD experiment.
QUaD has for the first time detected multiple acoustic peaks in
the {\it E}-mode polarization spectrum with high significance. 
Although QUaD-only parameter constraints are not competitive with previous results
for the standard 6-parameter $\Lambda$CDM cosmology,
they do allow meaningful polarization-only parameter analyses for the first time.

In a standard 6-parameter $\Lambda$CDM analysis we find the QUaD {\it TT}
power spectrum to be in good agreement with previous results.
However, the QUaD polarization data shows some tension with $\Lambda$CDM.
The origin of this $1-2\sigma$ tension remains unclear, and may 
point to new physics, residual systematics or simple random chance.
We also combine QUaD
with the five-year WMAP data set and the SDSS
Luminous Red Galaxies $4^{th}$ data release power spectrum, and extend
our analysis to constrain individual isocurvature mode fractions, constraining  cold dark
matter density, ${\alpha}_{\rm cdmi}<0.11$ ($95\%$ CL), neutrino density, ${\alpha}_{\rm
ndi}<0.26$ ($95\%$ CL), and neutrino velocity, ${\alpha}_{\rm nvi}<0.23$ ($95\%$ CL), modes.
Our analysis sets a benchmark for future polarization experiments.
\end{abstract}



\section{Introduction}
\label{sec:intro}
\setcounter{footnote}{0}

The anisotropies of the Cosmic Microwave Background (CMB)
radiation are among the most important tests of cosmology. The
large-angle Sachs-Wolfe effect, multiple acoustic oscillations and
the Silk damping tail in the temperature power spectrum have now
been confirmed by a range of experiments from the largest angular
scales down to angular scales of a few arcminutes
(\cite{dunkley08,reichardt08}). The full repository of CMB data
available, in conjunction with other cosmological observables,
such as data coming from the large-scale distribution of galaxies or
Supernova type Ia observations, are extremely well described by
the spatially flat $\Lambda$CDM cosmological model. 

A generic prediction of cosmology is that the CMB photons should
be polarized at the $10$\% level. The polarization field can be
decomposed into two components: primary even-parity, curl-free 
{\it E}-modes are generated at the last scattering surface by both scalar
and tensor metric perturbations (gravitational waves); primary
odd-parity {\it B}-modes are generated only by tensor perturbations due
to gravitational waves passing through the primordial plasma.
Secondary anisotropies in both the {\it E} and {\it B}-mode polarization arise
at the epoch of reionization, while {\it E} and {\it B}-modes are mixed by
gravitational lensing by intervening large-scale structure along
the line of sight (see eg \cite{art:huwhite97}). Observations of this 
linearly polarized component provide an
important consistency check of the standard model
and a detection of primordial
gravitational waves in the odd-parity {\it B}-mode on large angular
scales would be strong evidence for inflation.

After the DASI\footnote{DASI stands for ``Degree
Angular Scale Interferometer''.} experiment (\cite{kovac02})
made the first measurement of {\it E}-mode power, other
experiments have provided us with further measurements
at a wide range of angular scales
(\cite{barkats05,readhead04,montroy06,sievers05,page03,bischoff08,nolta08}).
Despite this we were still lacking precision measurements of its
power spectra down to arcminute scales, as we have for the
temperature. The {\it B}-mode polarization has not yet been detected and
only upper limits have been determined.

The QUaD\footnote{ QUaD stands for ``QUEST and DASI''. In turn,
QUEST is ``Q \& U Extragalactic Survey Telescope''. The two
experiments merged to become QUaD in 2003.} experiment is at the
forefront of this small-scale polarization quest, and after three
years of observations has delivered the highest resolution
{\it E}-mode spectrum and the tightest upper limits on the {\it B}-modes yet
measured. This is a significant improvement over the first season
of data results, previously reported by \cite{ade07}. In
particular, the sensitivity of QUaD has allowed us to see, for the
first time, four acoustic oscillations in the {\it E}-mode spectrum and
all significant oscillations in the {\it TE} spectrum to $\ell=2000$.
The overall consistency of peak phases and spacings between the
temperature and QUaD {\it EE} data was shown in~\cite{pryke08} 
(hereafter referred to as the "Power Spectra Paper"). 

In this paper we concentrate on using the QUaD temperature and
polarization power spectra to constrain the standard
cosmological model.
Using this baseline
model, we analyze the different contributions coming from each of
the QUaD spectra.
We also go beyond the standard LCDM model
using the QUaD data in combination with WMAP and SDSS 
to constrain an isocurvature contribution.


\section{Cosmological Parameter Estimation: Methodology}
\label{sec:param_estimation}

\subsection{Monte Carlo Markov Chain}
\label{sec:mcmc}

The Monte Carlo Markov Chain (MCMC) method
is a method designed to efficiently explore an unknown Probability
Distribution Function (PDF) by sequentially drawing samples from
it according to a proposal probability function
in our case the Metropolis algorithm (\cite{metropolis53} among others).
The ensemble of these
samples constitute a Markov Chain whose distribution corresponds
to that of the unknown PDF. We adopted the 
Gelman and Rubin {\it R}-statistic
to verify that our chains are properly
mixed and converged (\cite{gelman92,verde03}),
and compare sets of 4 chains of around 100 000 steps from which
we remove a burn-in period during which our criterium is
not met. We use at least 80 000
steps after burn-in.

The rate of convergence of the Markov chain is slowed down by
degeneracies between parameters, and the choice of the step size
in the Metropolis algorithm. Therefore we apply a standard
partial re-parametrization of the parameter space as suggested in
\cite{kosowsky02}. 
To further reduce the remaining degeneracies 
between parameters, we apply
a change of basis in parameter space described in~\cite{tegmark04}
which uses a covariance matrix to take account of all the
correlations between parameters. In the new basis,
the new parameters have zero average and unit variance. 

When we have a fair sample of the underlying distribution, the
MCMC method trivializes marginalization to a simple projection of
the points of the chain.
The mean marginalized value of each parameter 
will hereafter be called "the mean recovered model".
Note that some authors refer to the mean recovered model
as the ``best-fit model''.

To obtain constraints on the mean parameters values
one simply produces the 1D histograms of the chain values
for each parameter, and calculate confidence intervals
using the $p^{th}$ and $(1-p^{th})$ quantiles of the histograms 
as in \cite{verde03}.
Normally we use 68\% equivalent to a nominal 1-$\sigma$
constraint.
However if the constraint as defined above hits the prior boundary
on one end we instead choose the level
which contains 95\% of the total probability and quote an upper
(or lower) limit.

We shall also plot 2-D marginalized parameter distributions with 68\%
and 95\% contours estimated at $\Delta \ln L=-2.3$ and $-6.17$ from
the peak values. Assuming a Gaussian distribution we quote
$\chi^2$ values corresponding to our mean recovered model, and the
Probability To Exceed (PTE), $P(>\chi^2|\nu)$, which gives the
random probability to have found the measured value of $\chi^2$ or
greater by chance, for $\nu$ degrees of freedom. 


\subsection{The Likelihood and Nuisance Marginalization}
\label{sec:lik}

The likelihood for the measured $C_{\ell}$ bandpowers
is well approximated by a Gaussian distribution, given by
 \be
    P( \hat{C}_{b} \! \mid \! C^{\rm th}_{b}) \propto \exp \left[ -
    \frac{1}{2}
    \Delta C_b(p) M_{bb'}^{-1} \Delta C^\dag_{b'}(p) \right],
 \ee
where $\Delta C_b(p)=\hat{C}_{b} - C^{\rm th}_{b}(p)$,
$\hat{C}_{b}$ are the measured QUaD binned
$C_\ell$ bandpowers, $C^{\rm th}_b(p)$ are the theoretical power
spectra which depend on the cosmological parameters, $p$, and
which have been transformed to predictions of the binned spectra
by means of the experimental bandpower window function (BPWF), and
$M_{bb'}=\langle \Delta C_b \Delta C^\dag_{b'} \rangle$ is the
measured $\hat{C}_{b}$ bandpower covariance matrix (BPCM). 

The BPCM is estimated from an ensemble of simulations of the CMB sky, 
assuming a fixed fiducial $\Lambda$CDM, run through the
QUaD analysis pipeline (see e.g. \cite{brown05}).
In our analysis, this BPCM remains independent of the cosmological
parameters, but in principle we should vary it as we move
around parameter space.
This re-scaling of the sample variance with $C^{\rm th}$ is
equivalent to saying that 
$P(C^{\rm th}_{b}(\bp) \! \mid \! \hat{C}_{b}) $
is log-normally distributed, and in the case where noise is present,
offset log-normally distributed.
\citet{bond98} propose a transformation of the bandpowers
and BPCM which accounts for this effect and allows use of
a fixed BPCM.
We have tested the impact of this transformation 
on our full data set and
find that the difference in parameter estimates and uncertainties is
insignificant.

The Gaussian likelihood has
the added benefit that we can simplify the marginalization over
beam and calibration nuisance parameters. Indeed, in this case, one can directly apply 
an analytic marginalization scheme as in \cite{bridle02}. The resulting
marginalization results in extra terms added to the BPCM 
of our Gaussian likelihood that acts as
a source of extra noise. The likelihood function is then given by
 \be
    \ln L = -\frac{1}{2} \Delta C_b {M'}_{bb'}^{-1} \Delta C^\dag_{b'}
      - \frac{1}{2} {\rm Tr}  \ln M',
      \label{eq:lnL_nuis}
 \ee
where
 \be
        M'_{bb'} = M_{bb'} +  \sigma^2_{{\rm Cal}} C_b C^\dag_{b'}
        +2\sigma^2_{b}\delta \ell_b^2 C_b \ell_{b'}^2
        C^\dag_{b'},
 \ee
is the marginalized bandpower covariance matrix, where 
$\sigma^2_{{\rm Cal}}$ is the variance on the calibration, $\sigma_b=\theta_{FWHM}/\sqrt{8\ln{2}}$, 
$\theta_{FWHM}$ is the effective beam size, $\delta$ is the fractional beam
error and $\ell_b$ is the average multipole in a bin.

In assessing the goodness-of-fit of our
mean recovered models, and when comparing the 
measured $\hat{C}_{b}$ bandpowers
with the WMAP5 $\Lambda$CDM model, we shall use the
$\chi^2$-statistic, introduced previously in Section~\ref{sec:mcmc}, 
and which we define by
 \be
        \chi^2 =\Delta C_b {M'}_{bb'}^{-1} \Delta C^\dag_{b'},
 \ee
where we use the nuisance marginalized BPCM, ${M'}_{bb'}$, as 
defined in the previous equation.

\begin{figure*}[t]
\centering
\resizebox{14cm}{!}{\includegraphics{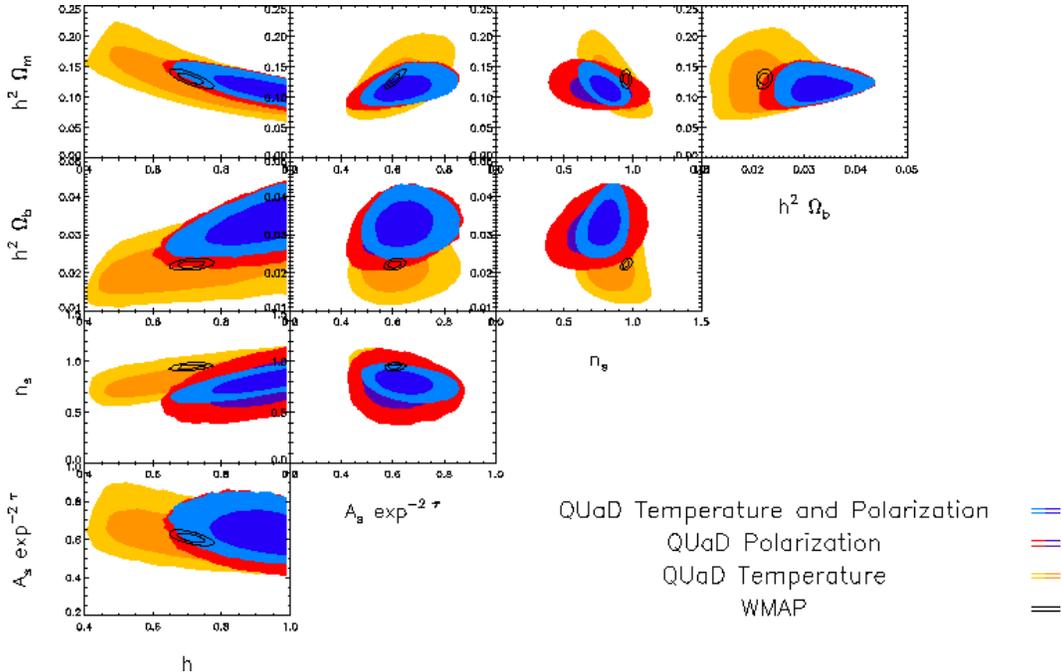}}
\caption{2-D projected basic parameter likelihood surfaces with
two-parameter 1- and 2-sigma contours for QUaD only constraints
using the {\it TT}/{\it TE}/{\it EE}/{\it BB} data set (TP: the blue contours), 
using the {\it TE}/{\it EE}/{\it BB} data set (P: the red \& magenta contours) and
using the {\it TT} spectrum (T: the yellow \& orange contours) versus
the WMAP5 constraints (black/empty contours). 
Pivot scale used is $k_p=0.05$~Mpc$^{-1}$. 
} 
\label{fig:2D_Q08_main}
\end{figure*}


\subsection{The Standard Cosmological Model}
\label{sec:stand_model}

We parameterize our flat $\Lambda$CDM cosmological model with
the following standard set of 6 cosmological parameters: the
Hubble constant, $H_0=100h$~${\rm kms^{-1}Mpc^{-1}}$; the physical matter
density, $\Omega_mh^2$; the physical baryon density, $\Omega_bh^2$; the
amplitude of scalar fluctuations, $A_s$; the scalar spectral
index, $n_s$; and the optical depth, $\tau$. 
When using QUaD data by itself, we present
the combination $A_se^{-2\tau}$ as our 
individual constraints on the degenerate parameters $A_s$ and $\tau$
are prior driven and thus biased, as explained
in Appendix~\ref{sec:sims_param}.
Initial conditions are taken to be purely
adiabatic with an initial power-law mass-density perturbation
spectrum. 
Due to the range of angular scales probed by QUaD, 
the pivot point we use when analyzing QUaD data by itself is 
$k_p = 0.05$~${\rm Mpc}^{-1}$ (note this is independent of $h$).
When comparing our QUaD results with WMAP we regenerate WMAP best-fit 
values based on this pivot value using our own pipeline, however when adding QUaD
data to other data sets for a combined analysis we revert to the WMAP preferred pivot scale
of $k_p = 0.002$~${\rm Mpc}^{-1}$.

To generate our theoretical spectra we use the publicly available
CAMB code (\cite{lewis00}), including the
effects of reionization, and gravitational lensing by foreground
structure. We impose the following flat priors in the likelihood
analysis: $0 \le \Omega_c h^2 \le 1$, $0 \le \Omega_b h^2 \le 1$,
$ 0.005 \le \theta \le 0.1$, $0 \le \tau \le 0.8$, $ 0 \le A_s
\le 2.5$ and $0 \le n_s \le 2$. 
The parameter $\theta$ is the
angular sound horizon, and $\Omega_c h^2$ is the 
physical cold dark matter density. Note that the
partial re-parametrization of the parameter space as suggested in
\cite{kosowsky02} introduces an implicit prior on
the $h$ parameter.


\section{Results: Basic 5-Parameter Constraints}

\subsection{QUaD Only Constraints}

\label{sec:results_quad}

The QUaD data set we use to constrain cosmological parameters are
the optimally combined spectra obtained from the 100 GHz, 150 GHz 
and frequency-cross temperature and polarization {\it E-} and {\it B-} 
power spectra, measured in 23 bandpowers over angular
multipoles from $200 < \ell <2000$
as described in the Power Spectra Paper.
We estimate a 10\% uncertainty 
on both the calibration (in power units) and the beam sizes, 
assuming an effective beam of 4.1 arcminutes~\footnote
{The data set is publicly available online at http://quad.uchicago.edu/quad}.

In parameter estimation we use
the diagonal and the first two off-diagonal terms of the BPCM 
for {\it TT}-{\it TT}, {\it TE}-{\it TE}, 
{\it EE}-{\it EE}, and {\it BB}-{\it BB} covariances,
but only the diagonal and first off-diagonal terms in the {\it TT}-{\it TE}
and {\it TE}-{\it EE} covariances.
This is motivated by the need to avoid
excessive noise in the off-diagonal terms of the BPCM, due to its
estimation from numerical simulations. 
We also ignore the covariance between {\it TT} and
{\it EE}, which is much smaller than the other terms.

We explore various sets of combinations of
the QUaD temperature and polarization data in order 
to understand the new information each
spectrum brings to parameter estimation. 
In Table~\ref{table:Q08_main} we present the mean recovered models.
Figure~\ref{fig:2D_Q08_main} shows the corresponding 2-D marginalized contour projections of
the likelihood in the 5-parameter space. 

\begin{deluxetable*}{c c c c c c c }
  \tablewidth{15cm}
  \tablecaption{
  Basic cosmological mean parameter constraints using QUaD bandpower spectra for various data combinations. \label{tab:param}} 
\scriptsize
\tablehead{ \colhead{Symbol} &\colhead{Q08 {\it TT}/{\it TE}/{\it EE}/{\it BB}} & \colhead{Q08 {\it TE}/{\it EE}/{\it BB}} & \colhead{ Q08
{\it TT}} & \colhead{Q08 {\it TE}} & \colhead{Q08 {\it EE}/{\it BB}} & \colhead{WMAP5}}
    \startdata
    $\Omega_bh^2$ & $0.0334\,\,_{-0.0040}^{+0.0039}$  & $0.0319 \pm 0.0046$ & $0.0242\,\,_{-0.0057}^{+0.0058}$ & $0.0398\pm0.0051$ & $0.0366\,\,_{-0.0161}^{+0.0159}$ &  $0.02261\,\,_{-0.00063}^{+0.00062}$\\
    $\Omega_mh^2$ & $0.119\,\,_{-0.015}^{+0.014}$ & $0.117 \pm 0.017$  &   $0.133\,\,_{-0.033}^{+0.035}$ & $0.149\pm0.025$  &   $0.155\,\,_{-0.035}^{+0.034}$ &  $0.1329\,\,_{-0.0065}^{+0.0064}$\\
    $h$ & $0.91 \pm 0.09$ & $0.90 \pm 0.10$ & $0.75\,\,_{-0.18}^{+0.17}$ & $0.87\pm0.11$ & $0.77\pm0.18$ & $0.717\,\,_{-0.027}^{+0.026}$ \\
    $A_se^{-2\tau}\,\,\tablenotemark{a}$ & $0.66 \pm 0.08$ & $0.63 \pm 0.09$ & $0.64\,\,_{-0.10}^{+0.09}$  & $0.63\pm0.13 $ & $0.79 \pm 0.21$ & $0.614\,\,_{-0.018}^{+0.017}$ \\
    $n_s\tablenotemark{a}$ & $0.809 \pm 0.078$ & $0.766 \pm 0.152$ &   $0.848\,\,_{-0.121}^{+0.117}$  & $1.337\,\,_{-0.254}^{+0.259}$ &   $0.534\,\,_{-0.161}^{+0.155}$ & $0.967\pm0.015$ \\ \\
    \hline
        \\
    $\chi^2(\nu)\tablenotemark{b}$ & $88.60\,(86)$ & $74.78\,(63)$ & $12.73\,(17)$   & $19.67\,(17)$  & $33.16\,(40)$   & \\
    $PTE$:  $P(\ge \chi^2|\nu)$   & $40.26$\% & $14.72$\% &  $ 75.38$\% &  $29.14$\% &  $76.94$\%  & \\
    $\chi^2(WMAP5|Q08) \tablenotemark{c}          $ & $108.63\,(92)$ & $86.99\,(69)$ & $14.48\,(23)$  & $31.44\,(23)$  & $41.62\,(46)$   & \\
    $PTE(WMAP5|Q08)  $   & $11.36$\% & $7.07$\% &  91.24\% &  $11.24$\% &  $65.60$\%  & \\ 
      \enddata
    \tablenotetext{a}{The pivot point for $A_s$ and $n_s$ is $k_p=0.05$~Mpc$^{-1}$ for both the QUaD data and WMAP5 data.}
     \tablenotetext{b}{$\chi^2$ for the 6-parameter mean recovered model against QUaD data, with the number of degrees of freedom in brackets.}
     \tablenotetext{c}{$\chi^2$ for WMAP5 mean recovered model given the QUaD data set, with the number of degrees of freedom in brackets.}
\label{table:Q08_main}
\end{deluxetable*}

All of the statistics shown verify that the QUaD {\it TT} temperature power spectrum is
compatible with the results from WMAP5. This is a non-trivial
test, since the overlap of scales measured by QUaD and WMAP5 is only in the
range $\ell \approx 200$ to $\ell \approx 950$, while the QUaD data extends to
$\ell \approx 2000$ with good signal-to-noise. 

However the analysis involving the polarization spectra are in less good agreement,
yielding a high value of the baryon content. Indeed for the {\it TE}/{\it EE}/{\it BB} combination
we have $\Omega_b h^2=0.0319 \pm 0.0046$, compared to
$\Omega_b h^2=0.02261 \pm 0.00062$ from WMAP5. 
The $\chi ^2$ of the WMAP5 best-fit model for this data set
has a PTE of $7\%$ indicating a modest degree of tension.
The spectrum responsible
for this behaviour seems to be the {\it TE} spectrum.

\begin{figure}[t]
  \centering
\resizebox{6cm}{!}{\includegraphics{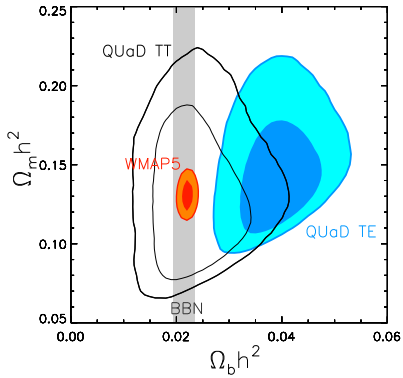}}
\caption{2-D marginalized contours of the parameters $\Omega_b h^2$
versus $\Omega_mh^2$ obtained from QUaD {\it TE} data only. Also plotted
are the contours from QUaD {\it TT} data only, the
results from WMAP5, and the BBN constraint (\cite{art:KirkmanEtAl03}).
}
\label{fig:2D_Q08_TE_ombhh_ommhh}
\end{figure}

\begin{figure}th]
\centering
\resizebox{8cm}{!}{\includegraphics{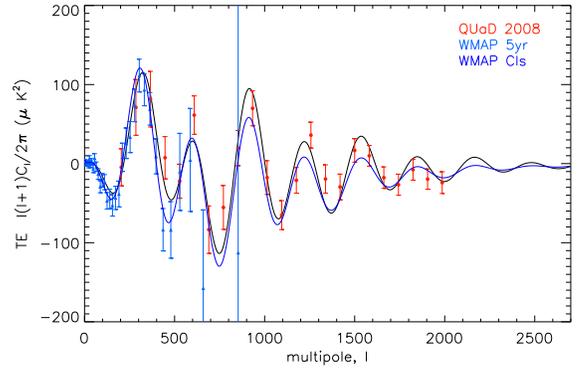}}
\caption{Plot of the QUaD TE data bandpower spectrum (in red)
versus the QUaD TE mean recovered model (in solid black)
together with WMAP5 best-fit model
(blue). For the TE mean recovered model, we assumed the WMAP5 best-fit value for the
optical depth ($\tau=0.087$), which corresponds to an Amplitude of $A_s=0.75$, given 
our $A_se^{-2\tau}$ constraint.} 
\label{fig:cl_Q08_TE_vs_wmap}
\end{figure}

Clearly the {\it TE} only constraints
are weak and most parameters are prior driven, but
surprisingly we obtain constraints on $\Omega_bh^2$ and
$\Omega_mh^2$ that are not influenced by their choice of priors.
Figure~\ref{fig:2D_Q08_TE_ombhh_ommhh} shows the 2-D projected
likelihood surface for the ($\Omega_bh^2$,$\Omega_mh^2$) parameter
space. To illustrate the difference with the 
{\it TT} only contours we overplot them.
In addition we show the results from WMAP5 and the Big 
Bang Nucleosynthesis (BBN)
constraint of $\Omega_bh^2 = 0.0214 \pm 0.002$ from \cite{art:KirkmanEtAl03}.
We also show in Figure~\ref{fig:cl_Q08_TE_vs_wmap} the {\it TE} QUaD bandpower
spectrum versus its mean recovered model and the WMAP5 full data set
best-fit model. This figure visually illustrates the differences of
the mean recovered models between the two data sets in terms
of height and location of the peaks.  The main
reason for the higher baryon density parameter seems to
be larger acoustic oscillations at higher multipole in {\it TE}, as well as
a shift to higher multipoles of the peaks, resulting in a slight degeneracy with $h$, which may 
explain its high value. The origin of this source of tension is unclear, but
could be due to a new physical mechanism, residual systematics or random chance.

Another interesting result comes from the  {\it EE} and {\it BB} spectra. 
As expected they provide very little information on parameters, and the 
$\chi ^2$ of the WMAP5 best-fit model for this data set
has an acceptable PTE of $65.6\%$. They do however have an
unusual feature; the preferred range of scalar spectral index
values is low ($n_s = 0.534\,\,_{-0.161}^{+0.155} $). So although
{\it TT} and {\it TE} share the majority of the constraining power, the {\it EE} and
{\it BB} spectra exert an influence in combination with the remaining
spectra by restricting the $n_s$-range to low values.

\begin{figure}[h]
\resizebox{\columnwidth}{!}{\includegraphics{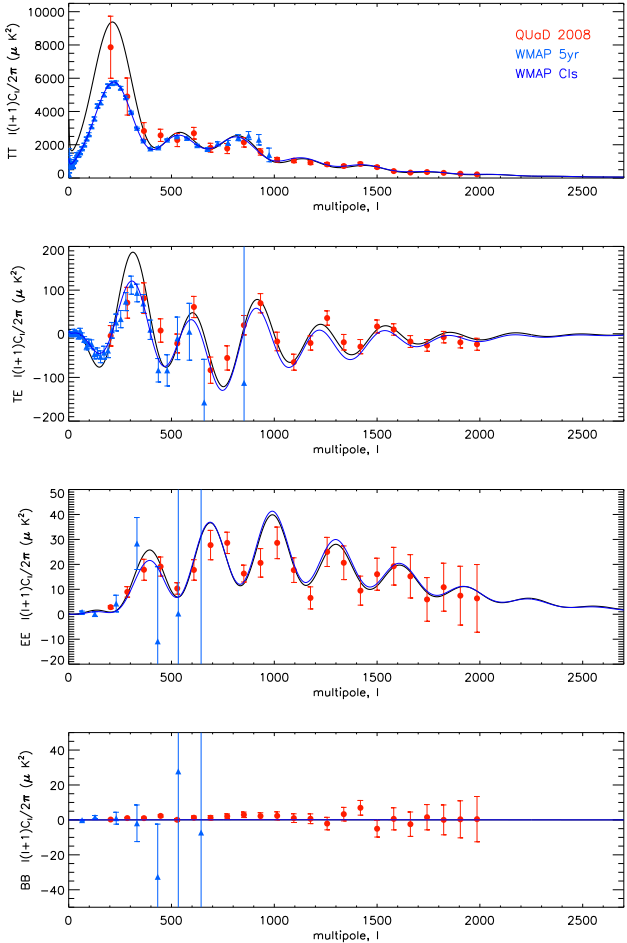}}
\caption{The QUaD power spectra used in this analysis (red points) for {\it TT}, {\it TE}, 
{\it EE} and {\it BB} (top to bottom). The
blue data points are WMAP 5-year power spectra data. The blue line
shows the basic WMAP5 best-fit model as defined in \cite{dunkley08}, while
the black solid line shows our {\it TT}/{\it TE}/{\it EE}/{\it BB} mean recovered model with values
given in Table~\ref{table:Q08_main}.  We assumed the WMAP5 best-fit value for the
optical depth ($\tau=0.087$), which corresponds to an Amplitude of $A_s=0.79$, given 
our $A_se^{-2\tau}$ constraint.  } 
\label{fig:spec_comb_TP_bf}
\end{figure}

If we combine all the spectra together then the polarization data dominate the constraints.
The majority of parameters are consistent with
the WMAP5 results, but the spectral index $n_s$ is lower,
influenced by the {\it EE}/{\it BB} contribution, and
the Hubble parameter $h$ and $\Omega_b h^2$ are
higher, driven by the polarization data in particular the {\it TE}
spectrum. Compared to the BBN value $\Omega_b h^2$
is almost 3-$\sigma$ away. Figure~\ref{fig:spec_comb_TP_bf}
shows a comparison between our {\it TT}/{\it TE}/{\it EE}/{\it BB} 
data and models and the WMAP5 data and best-fit model. 
The $\chi^2$ of the WMAP5 best-fit model indicates that there is
an $11.36 \%$ chance that the combined QUaD spectra are
a realization of this model.

\begin{deluxetable*}{ c c c c}
  \tablewidth{15cm}
  \centering
  \tablecaption{
    Basic mean parameters for QUaD {\it TT}/{\it TE}/{\it EE}/{\it BB}, SDSS LRG and WMAP5 data.
\label{tab:param}} \tablehead{ \colhead{Symbol} &
\colhead{Q08 {\it TT}/{\it TE}/{\it EE}/{\it BB}+WMAP5} & \colhead{Q08
{\it TT}/{\it TE}/{\it EE}/{\it BB}+WMAP5+SDSS} & \colhead{WMAP5} }
    \startdata
    $\Omega_bh^2$ & $0.02233\,\,_{-0.00057}^{+0.00055}$ & $0.02235\,\,_{-0.00058}^{+0.00051}$  & $0.02259\,\,_{-0.00062}^{+0.00061}$\\
    $\Omega_mh^2$ & $0.1266\pm0.0060$ & $0.1266\,\,_{-0.0039}^{+0.0038}$ & $0.1329\pm0.0065$ \\
    $h$ & $0.733\pm0.027$ & $0.731\pm0.019$  & $0.715\,\,_{-0.026}^{+0.027}$ \\
    $\tau$ & $  0.087\pm0.017   $  & $ 0.087\pm0.016 $  & $ 0.087\pm0.017 $ \\
    $A_s$\tablenotemark{a} & $0.805\pm0.038$ &$0.806\,\,_{-0.033}^{+0.032}$  & $0.816\pm0.039$  \\
    $n_s$\tablenotemark{a} & $0.960\,\,_{-0.013}^{+0.014}$ & $0.960\,\,_{-0.012}^{+0.014}$ & $0.966\,\,_{-0.015}^{+0.014}$ \\
    \enddata
    \tablenotetext{a}{The pivot point for $A_s$ and $n_s$ is $k_p=0.002$~Mpc$^{-1}$ for QUaD, WMAP5 and SDSS LRG data.}
\label{table:Q08_W5}
\end{deluxetable*}


\subsection{Combining QUaD with Other Data Sets}
\label{sec:results_quad_wmap}

In this section we will add to the QUaD spectra
the WMAP 5-year {\it TT}, {\it TE}, {\it EE} and {\it BB} data set. 
We use the WMAP5 likelihood code, publicly
available on the LAMBDA 
website\footnote{LAMBDA website: http://lambda.gsfc.nasa.gov/}, and their
methodology (\cite{dunkley08}), but do not include the Sunyaev
Zel'dovich (SZ) marginalization.
We shall assume the two
data sets are independent, as the QUaD data only covers a small
fraction of the WMAP5 sky, and the overlap in multipole range
is only partial. 

We will further add large-scale structure data from the SDSS
Luminous Red Galaxies (LRG) fourth data release using publicly available
likelihood code, measurements and window functions
(\cite{tegmark06}). Results from the SDSS LRG and the main 
SDSS galaxy samples are consistent, but
the former provides higher signal-to-noise ratio.
We use the SDSS LRG matter power spectrum over
wavenumbers $0.07h{\rm Mpc}^{-1} < k_{SDSS} < 0.2 h {\rm Mpc}^{-1}$
 so that we do not have to consider any
nonlinear correction. We marginalize over the amplitude of the
galaxy power spectrum which removes any dependence on the galaxy
bias parameter, $b_{g}$, and linear redshift-space distortion.

As can be seen in Table~\ref{table:Q08_W5},
the QUaD {\it TT}/{\it TE}/{\it EE}/{\it BB} power spectra have little impact on the
baseline 6-parameter mean parameter fit when combined with WMAP5. This is perhaps
unsurprising, given the accuracy of the WMAP5 measurement of the
first acoustic peak in the {\it TT} spectrum, and its
low-$\ell$ power in {\it TT} and {\it TE}. The impact it does have is to
tighten the error bars on parameters determined from the relative
heights of acoustic peaks, i.e.\ on the baryon density, $\Omega_bh^2$,
and the matter density, $\Omega_mh^2$, as QUaD data adds
a substantial amount of well-defined peak information at
high-$\ell$. 

When we combine the SDSS LRG and WMAP5 data with the QUaD data we see an
improvement compared to the QUaD and WMAP5 combination, as
expected. This improvement is mostly due to the extra constraining
power on $\Omega_mh^2$ and $\Omega_bh^2$ coming from the
break-scale in the SDSS galaxy power spectrum, and the baryon
acoustic oscillations. However the QUaD data still reduces
the uncertainty on $\Omega_bh^2$.

\begin{deluxetable*}{ c c c c}
  \tablewidth{15cm}
  \centering
  \tablecaption{
    CDM Isocurvature mean parameter constraints for QUaD {\it TT}/{\it TE}/{\it EE}/{\it BB}, WMAP5 and SDSS LRG data.
\label{tab:param}} \tablehead{ \colhead{Symbol} & \colhead{Q08 {\it TT}/{\it TE}/{\it EE}/{\it BB}+WMAP5}
& \colhead{Q08 {\it TT}/{\it TE}/{\it EE}/{\it BB}+WMAP5+SDSS} & \colhead{WMAP5} }
    \startdata
    $\Omega_bh^2$ & $0.02312\,\,_{-0.00081}^{+0.00080}$ & $0.02280\pm0.00070$  & $0.02362\,\,_{-0.00094}^{+0.00096}$\\
    $\Omega_mh^2$ & $0.1214\,\,_{-0.0068}^{+0.0067}$ & $0.1256\,\,_{-0.0037}^{+0.0038}$ & $0.1279\,\,_{-0.0071}^{+0.0072}$ \\
    $h$ & $0.773\pm0.039$ & $0.746\pm0.022$  & $0.759\pm0.041$ \\
    $\tau$ & $  0.087\pm0.017  $  & $ 0.084\,\,_{-0.016}^{+0.016} $  & $ 0.087\pm0.017 $ \\
    $A_s$\tablenotemark{a} & $0.786\pm0.037$ & $0.797\,\,_{-0.030}^{+0.031}$  & $0.789\pm0.038$  \\
    $n_s$\tablenotemark{a} & $0.987\pm0.023$ & $0.976\,\,_{-0.017}^{+0.018}$ & $0.998\,\,_{-0.026}^{+0.027}$ \\
    ${\alpha}_{\rm cdmi}$ (95 \% cl) & $  < 0.19  $  &  $< 0.11$ & $ < 0.21 $ \\
    \enddata
   \tablenotetext{a}{The pivot point for $A_s$ and $n_s$ for all isocurvature constraints is $k_p=0.002$~Mpc$^{-1}$ for QUaD, WMAP and SDSS data.}
\label{table:Q08_W5_SDSS_cdmi}
\end{deluxetable*}

\begin{deluxetable*}{ c c c c}
  \tablewidth{15cm}
  \centering
  \tablecaption{
    NDI Isocurvature mean parameter constraints for QUaD {\it TT}/{\it TE}/{\it EE}/{\it BB}, WMAP5 and SDSS LRG data.
\label{tab:param}} \tablehead{ \colhead{Symbol} & \colhead{Q08 {\it TT}/{\it TE}/{\it EE}/{\it BB}+WMAP5}
& \colhead{Q08 {\it TT}/{\it TE}/{\it EE}/{\it BB}+WMAP5+SDSS} & \colhead{WMAP5} }
    \startdata
    $\Omega_bh^2$ & $0.02370\pm0.00110$ & $0.02300\pm0.00080$  & $0.02410\,\,_{-0.00120}^{+0.00130}$\\
    $\Omega_mh^2$ & $0.1200\,\,_{-0.0072}^{+0.0073}$ & $0.1261\pm0.0039$ & $0.1270\,\,_{-0.0077}^{+0.0075}$ \\
    $h$ & $0.800 \pm 0.030$ & $0.751\,\,_{-0.025}^{+0.027}$  & $0.780\,\,_{-0.055}^{+0.058}$ \\
    $\tau$ & $  0.090\pm0.017   $  & $ 0.085\pm0.016 $  & $ 0.090\pm0.017 $ \\
    $A_s$ & $0.869\pm0.058$ & $0.855\,\,_{-0.051}^{+0.053}$  & $0.872\pm0.059$  \\
    $n_s$ & $0.995\,\,_{-0.0252}^{+0.0271}$ & $0.976\,\,_{-0.017}^{+0.020}$ & $1.003\,\,_{-0.030}^{+0.031}$ \\
    ${\alpha}_{\rm ndi}$  (95 \% cl) & $  < 0.37  $  & $ < 0.26 $ & $ < 0.38 $ \\
    \enddata
\label{table:Q08_W5_SDSS_ndi}
\end{deluxetable*}

\begin{deluxetable*}{ c c c c}[t]
  \tablewidth{15cm}
  \centering
  \tablecaption{
    NVI Isocurvature mean parameter constraints for QUaD {\it TT}/{\it TE}/{\it EE}/{\it BB}, WMAP5 and SDSS LRG data.
\label{tab:param}} \tablehead{ \colhead{Symbol} & \colhead{Q08 {\it TT}/{\it TE}/{\it EE}/{\it BB}+WMAP5}
& \colhead{Q08 {\it TT}/{\it TE}/{\it EE}/{\it BB}+WMAP5+SDSS} & \colhead{WMAP5} }
    \startdata
    $\Omega_bh^2$ & $0.02350\pm0.00090$ & $0.02339\,\,_{-0.00070}^{+0.00080}$  & $0.02390\pm0.00100$\\
    $\Omega_mh^2$ & $0.1260\pm0.0060$ & $0.1277\pm0.0040$ & $0.1330\pm0.0063$ \\
    $h$ & $0.745\pm0.029$ & $0.734\pm0.018 $  & $0.728\,\,_{-0.028}^{+0.027}$ \\
    $\tau$ & $  0.088\pm0.017  $  & $ 0.087\,\,_{-0.015}^{+0.016} $  & $ 0.089\pm0.018 $ \\
    $A_s$ & $0.851\,\,_{-0.047}^{+0.048}$ & $0.854\,\,_{-0.045}^{+0.044}$  & $0.859\pm0.048$  \\
    $n_s$ & $0.980\,\,_{-0.018}^{+0.017}$ & $0.978\,\,_{-0.014}^{+0.015}$ & $0.988\,\,_{-0.018}^{+0.019}$ \\
    ${\alpha}_{\rm nvi}  (95 \% cl)$ & $  < 0.27  $  & $ < 0.23 $ & $ < 0.27 $ \\
    \enddata
\label{table:Q08_W5_SDSS_nvi}
\end{deluxetable*}


\section{Beyond The standard 6-Parameter Model: Isocurvature Modes}
\label{sec:beyond_std_model}

Theoretical predictions of isocurvature modes and their evolution,
and the role of CMB polarization observations in constraining them,
has been an active field over the past few years 
(\cite{kawasaki07,Keskitalo:2006qv,Bean:2006qz,Beltran:2004uv, Moodley:2004nz} among many others).
Pure isocurvature perturbations have been ruled out
(\cite{Stompor:1995py,Langlois:2000ar,Enqvist:2000hp,Amendola:2001ni}) although the presence of a subdominant
isocurvature fraction has been claimed (\cite{Keskitalo:2006qv}).
Observationally, isocurvature modes have a phase difference from
adiabatic modes, which provides a distinct signature. 

We can completely characterize the primordial perturbations by one
adiabatic and several isocurvature components. The adiabatic
component is given by the associated curvature perturbation
$\mathcal{R}$ corresponding to an initial overdensity
$\delta=\delta\rho/\rho$. The non-adiabatic components are given
by entropy perturbations
$\mathcal{S}_x=\delta_x-(3/4)\delta_\gamma$ between photons and a
different species, $x$. These correspond to four possible
nondecaying isocurvature modes: baryon density, cold dark matter density 
(cdmi), neutrino density (ndi) and neutrino velocity (nvi). \cite{Bucher2000} 
have presented a thorough analysis of these components.

We parameterize the contribution of adiabatic and isocurvature
modes to the total temperature and polarization power spectra by
 \be
    C^X_\ell=A_s^{2} \big[(1-\alpha) \hat{C}_\ell^{X,{\rm Ad}}+\alpha
    \hat{C}_\ell^{X,{\rm Iso}}\big],
 \ee
where $\alpha$ is the isocurvature fraction. The adiabatic
spectra, $C_\ell^{X,{\rm Ad}}$, and the isocurvature spectra,
$C_\ell^{X,{\rm Iso}}$, are defined with unit amplitude and the 
same spectral index. In this
analysis we shall assume there is no correlation between adiabatic
and isocurvature modes, and will constrain one 
isocurvature mode at a time.  Also we
do not present results for the baryon density isocurvature mode as
these have a very similar signature to the cold dark matter mode. 

We analyze the QUaD {\it TT}/{\it TE}/{\it EE}/{\it BB} power spectra combined with WMAP5, and combined 
with the WMAP5 plus the SDSS LRG data. 
The shape of the galaxy power spectrum is sensitive to an isocurvature
contribution, and has been used in the past to improve on 
isocurvature constraints (e.g.~\cite{Beltran:2004uv,Beltran:2005gr,Crotty:2003rz}).
The results we obtain are given in
Table~\ref{table:Q08_W5_SDSS_cdmi} for the cdmi
mode, in Table~\ref{table:Q08_W5_SDSS_ndi} for the
ndi mode and
in Table~\ref{table:Q08_W5_SDSS_nvi} for the nvi mode.

Our analysis shows a small improvement in the isocurvature 
cold dark matter constraint
when we add the QUaD to the WMAP5 data, from $\alpha_{\rm
cdmi}<0.21$ to $\alpha_{\rm cdmi}<0.19$ ($95\%$
confidence limits). In addition we find an improvement in the
$\Omega_bh^2$ and $\Omega_mh^2$ constraints. There is a similar
improvement for the neutrino density isocurvature constraints: we
go from $\alpha_{\rm ndi}<0.38$ to $\alpha_{\rm ndi}<0.37$. For
the neutrino velocity isocurvature modes there is no improvement,
the constraint staying at $\alpha_{\rm nvi}<0.27$.

We can further improve on these results by adding the SDSS LRG
data. When we do this the 
cold dark matter isocurvature constraint becomes $\alpha_{\rm cdmi}<0.11$ (95 \% CL). 
The largest improvement is for the neutrino density isocurvature mode, $\alpha_{\rm
ndi}<0.26$, while the smallest improvement is for the neutrino
velocity isocurvature mode, $\alpha_{\rm nvi}<0.23$.


\section{Summary and Conclusions}
\label{sec:conclusions}

We have presented a standard cosmological parameter constraint analysis, 
and its extension to include isocurvature modes,
using QUaD {\it TT}, {\it TE}, {\it EE} and {\it BB} 
bandpower spectra.
This is the first CMB experiment to detect with confidence the
acoustic oscillations in the {\it EE} spectrum and so the first to
be able to provide significant constraints on cosmological parameters from
individual CMB polarization spectra ---
in fact our polarization-only constraints are superior to our
temperature-only constraints.
In combination with the WMAP5 data set
QUaD offers a small improvement in the constraints on the baryon and matter densities.

We find our QUaD temperature data is in good agreement with the results
from WMAP5, which is a non trivial test of LCDM
as the QUaD data extends to $\ell\approx2000$ with good signal-to-noise.
However, our polarization ({\it TE}, {\it EE} and {\it BB}) data 
is in less good agreement yielding a higher 
baryon density value of $\Omega_b h^2=0.0319
\pm 0.0046$, compared with $\Omega_b h^2=0.0242\,\,_{-0.0057}^{+0.0058}$ from our {\it TT} data 
and $0.02261\,\,_{-0.00063}^{+0.00062}$ from our
re-analysis of WMAP5.
A $\chi^2$ test shows there is a
$7\%$ probability of the QUaD polarization results arising by chance,
assuming the WMAP5 $\Lambda$CDM model is correct. Although not of high significance,
this modest level of tension, that seems to originate from the {\it TE} spectrum, 
could be due to new physics in polarization, residual systematics in the data, or 
random chance.
It will be interesting to see
if this trend continues in future polarization experiments.

We also investigate isocurvature cold dark matter density,
neutrino density and neutrino velocity modes.
We find QUaD provides a marginal improvement
on the fractional cold dark matter density
mode parameter, ${\alpha}_{\rm cdmi}$, from $<0.21$ for WMAP5 alone to
$<0.19$.


\acknowledgements

QUaD is funded by the National Science Foundation in the USA,
through grants AST-0096778, ANT-0338138, ANT-0338335 \&
ANT-0338238, by the UK Science and Technology Facilities Council (STFC)
and its predecessor the Particle Physics and Astronomy Research Council
(PPARC), and by the Science Foundation Ireland.
JRH acknowledges the
support of an NSF Graduate Research Fellowship, a Stanford
Graduate Fellowship and a NASA Postdoctoral Fellowship. 
CP and JEC acknowledge partial support from
the Kavli Institute for Cosmological Physics through the grant NSF
PHY-0114422.  EYW acknowledges receipt of an NDSEG fellowship. YM
acknowledges support from a SUPA Prize studentship. PGC
acknowledges funding from the Funda\c{c}\~{a}o para a Ci\^{e}ncia e a Tecnologia. 
MZ acknowledges support from a NASA Postdoctoral Fellowship.
This work has made
use of the resources provided by the Edinburgh Compute and Data
Facility (ECDF) that is partially supported by the eDIKT
initiative. We also thank Licia Verde and Joanna Dunkley for useful
discussion.
 

\appendix

\section{Simulating Parameter Estimation}
\label{sec:sims_param}

\begin{figure*}[t]
\centering
\resizebox{11cm}{!}{\includegraphics{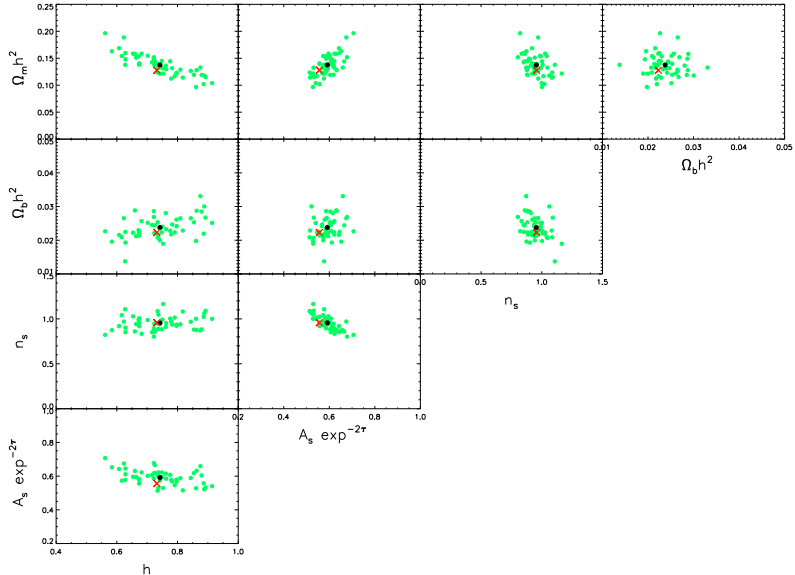}}
\caption{2D marginalized plot, showing the scattered values (green) of the 5 parameters mean recovered
basic cosmological model obtained from 50 simulations of QUaD 
{\it TT}, {\it TE}, {\it EE} and {\it BB} data
generated from the WMAP3 best-fit model (in red) from table 2 of~\cite{spergel06}.
The mean over the 50 simulations is shown as
a blue point.} 
\label{fig:sims_Atau}
\end{figure*}

We test our MCMC pipeline by running it on
a set of 50 simulations of the QUaD {\it TT}, {\it TE}, {\it EE} and {\it BB} bandpower spectra.
These were generated by simulating the signal and noise properties of the
time-ordered data and passing these through the QUaD pipeline in
the same way as the data (see the Power Spectra Paper for details).
The input cosmological model for these simulations was the WMAP3
mean recovered model (see Table 2 of~\cite{spergel06}).
The scatter in the values of the mean recovered model obtained from each one
of the 50 simulations can be seen in Figure~\ref{fig:sims_Atau}.
We also overplot the average values for each parameter 
calculated from the 50 simulations (see blue points).
We have verified that the scatter in the simulated mean parameter
results is close to the size of the contours produced by our
MCMC code when using real data, indicating that our code
accurately estimates the parameter uncertainties.

We can also compare the mean parameter values and the scatter about
them with the input WMAP3 best-fit model
(red crosses).
The average over the simulations
closely matches the input model indicating that our parameters are not biased.
If constrained independently, the scalar amplitude, $A_s$, and the optical depth, $\tau$,
parameters are biased, their values being systematically
higher than the input values.
This is due to the
combination of the large degeneracy between the amplitude, $A_s$,
and optical depth, $\tau$, and the parameter priors.
To break this degeneracy requires large-scale polarization
measurements probing the re-ionization bumps at lower $\ell$-modes.
As can be seen in the Figure, this problem can be avoided if we combine $A_s$ and $\tau$
into the parameter $A_se^{-2\tau}$ along the line of 
degeneracy, which is the approach followed in Section~\ref{sec:results_quad}.


\bibliographystyle{apj}

\bibliography{cosmo}

\end{document}